\documentclass[12pt]{JHEP3}

\usepackage{cite}
\usepackage{enumerate}
\usepackage{amsbsy}
\usepackage[dvips]{graphicx}
\usepackage{graphics}

\newcommand{\be}{\begin{equation}} \newcommand{\ee}{\end{equation}}
\newcommand{\bea}{\begin{eqnarray}} \newcommand{\eea}{\end{eqnarray}}
\newcommand{\el}{\nonumber \\}
\newcommand{\re}[1]{(\ref{#1})}

\newcommand{\pat}{\partial}

\renewcommand{\sec}[1]{section \ref{#1}}

\newcommand{\brt}[1]{[#1]}
\newcommand{\para}{\paragraph}

\renewcommand{\a}{\alpha}
\renewcommand{\b}{\beta}
\renewcommand{\c}{\gamma}
\renewcommand{\d}{\delta}
\newcommand{\e}{\epsilon}
\renewcommand{\l}{\lambda}

\newcommand{\GN}{G_{\mathrm{N}}}
\newcommand{\ha}{\frac{1}{2}}

\newcommand{\rmd}{\mathrm{d}}

\newcommand{\hi}{{\hat{i}}}
\newcommand{\hj}{{\hat{j}}}

\newcommand{\nonum}{\\}
\newcommand{\etal} {et al.}

\newcommand{\adot}{\dot{a}}
\newcommand{\addot}{\ddot{a}}

\newcommand{\Psidot}{\dot{\Psi}}

\newcommand{\acc}{\frac{\addot}{a}}

\newcommand{\sigmat}{\tilde{\sigma}}

\newcommand{\path}{\hat{\nabla}}

\newcommand{\tb}{\bar{t}}
\newcommand{\pt}{\tilde{\pat}}

\newcommand{\av}[1]{\langle{#1}\rangle}
\newcommand{\sQ}{\mathcal{Q}}
\newcommand{\sR}{{^{(3)}R}}

\newcommand{\sN}{\mathcal{N}}

\newcommand{\sO}{\mathcal{O}}

\newcommand{\PRD}[1]{{\it Phys. Rev.} {\bf D#1}}

\newcommand{\PRL}[1]{{\it Phys. Rev. Lett.} {\bf #1}}

\newcommand{\PLA}[1]{{\it Phys. Lett.} {\bf A#1}}

\newcommand{\MNRAS}[1]{{\it Mon. Not. Roy. Astron. Soc.} {\bf #1}}
\newcommand{\APJ}[1]{{\it Astrophys. J.} {\bf #1}}

\newcommand{\CQG}[1]{{\it Class. Quant. Grav.} {\bf #1}}
\newcommand{\GRG}[1]{{\it Gen. Rel. Grav.} {\bf #1}}
\newcommand{\AaA}[1]{{\it Astron. \& Astrophys.} {\bf #1}}
\newcommand{\PROG}[1]{{\it Prog. Theor. Phys.} {\bf #1}}

\newcommand{\IJMPD}[1]{{\it Int. J. Mod. Phys.} {\bf D#1}}

\title{Light propagation and the average expansion rate in near-FRW universes}

\author{Syksy R\"{a}s\"{a}nen \\

University of Helsinki, Department of Physics \\
and Helsinki Institute of Physics \\
P.O. Box 64, FIN-00014 University of Helsinki, Finland \\

\email{syksy {\it dot} rasanen {\it at} iki {\it dot} fi}}

\abstract{
We consider universes that are close to Friedmann-Robertson-Walker
in the sense that metric perturbations, their time derivatives
and first spatial derivatives are small, but second
spatial derivatives are not constrained.
We show that if we in addition assume that the observer four-velocity is
close to its background value and close to the four-velocity
which defines the hypersurface of averaging, the redshift and
the average expansion rate remain close to the FRW case.
However, this is not true for the angular diameter distance.
The four-velocity assumption implies certain conditions on
second derivatives of the metric and/or the matter content.
}

\preprint{HIP-2011-19/TH}

\begin{document}
  
\setcounter{tocdepth}{2}

\setcounter{secnumdepth}{3}

\section{Introduction} \label{sec:intro}

%\para{Backreaction and perturbativity.}

The matter-dominated homogeneous and isotropic cosmological model
based on general relativity disagrees with observations at late times.
The observed angular diameter distance to the last scattering surface
at redshift 1090 is a factor of 1.4--1.9 longer (keeping the Hubble
constant fixed) \cite{Vonlanthen:2010, Hubble},
and the expansion rate is larger by a factor of 1.2--1.7
(keeping the age of the universe fixed, i.e. $H_0 t_0\approx0.8\ldots1.1$)
\cite{Krauss:2003, Hubble}
or by a factor of 1.6--2.2 (keeping the matter density fixed,
i.e. $\Omega_{m0}\approx0.2\ldots0.4$)
\cite{Peebles:2004, Vonlanthen:2010, Hubble}.
The usual remedy is to either include exotic matter with negative
pressure or modify the law of gravity. However,
homogeneous and isotropic models do not
include the effect of non-linear structures on the expansion
of the universe and on light propagation, and the factor two
failure of the predictions of the matter-dominated model could
be related to this shortcoming
\cite{Buchert:2000, Tatekawa:2001, Wetterich:2001, Schwarz:2002, Rasanen}.

The effect of inhomogeneities on the average expansion rate is called
backreaction \cite{fitting, Buchert:1995, Buchert:1999}; see
\cite{Ellis:2005, Rasanen:2006b, Buchert:2007, newrevs} for reviews.
It has been shown in toy models that non-linearities
can lead to faster expansion, even acceleration, for dust matter
\cite{Rasanen:2006b, Chuang:2005, Paranjape:2006a, Kai:2006,
Rasanen:2006a, Paranjape:2009}.
In a semi-realistic model, the observed timescale and the order
of magnitude of the change in the expansion rate emerge from the
physics of structure formation \cite{Rasanen:2008a, peakrevs},
but there is no fully realistic calculation yet.
If backreaction is significant (and the universe is statistically
homogeneous and isotropic with a homogeneity scale smaller than
the horizon), this has to be due to non-Newtonian aspects of gravity
\cite{Buchert:1995, Buchert:1999, Notari:2005, Kolb:2005a, Kolb:2005b, Rasanen:2006b, Buchert:2007, Rasanen:2008a, Buchert:2008, Rasanen:2010a}
which are related to the difference between Newtonian gravity and the
weak field, small velocity limit of general relativity
\cite{Ellis:1971, Ehlers:1991, Ellis:1994, Senovilla:1997, vanElst:1998, Ehlers:1998, Ehlers:1999, Szekeres}.

The magnitude of the effect in the real universe remains an open
question. It has been argued that backreaction is small because the
universe is close to a homogeneous and isotropic
Friedmann-Robertson-Walker (FRW) model at all times.
However, we should be specific ab
out what is meant with
the statement that the universe is close to FRW.
Smallness of metric perturbations does not preclude large
deviations in the Riemann tensor, because the latter involves
second derivatives of the metric, and the variation
of a function may be rapid (with regard to some relevant
scale) even though its amplitude is small. This is the case in
cosmology when density perturbations enter the non-linear regime.

The argument involves two separate questions. First, can the
universe can be described with a metric which is perturbatively
close to the FRW metric even after density perturbations are
non-linear?
Second, does smallness of metric perturbations imply that the
average expansion rate, the redshift and the angular diameter
distance remain close to their unperturbed values?

The second question has been studied in many papers
with regard to the average expansion rate
\cite{Futamase, Seljak, Russ:1996, Wetterich:2001, Rasanen, Kolb:2004, Siegel:2005, Ishibashi:2005, Rasanen:2010a, Baumann:2010}
(see \cite{Rasanen:2010a} for further discussion and references).
However, almost all studies have been restricted to first or second
order perturbation theory and/or have had other shortcomings
\cite{Rasanen:2010a}\footnote{In \cite{Baumann:2010}, the average is
taken over the background FRW volume as opposed to the physical volume,
so the central issue of the non-commutativity of time derivatives and
averaging is missing.}.
A notable exception is \cite{Green:2010}, which considers
a new perturbative formalism adapted to the cosmological
situation where ordinary perturbation theory is not applicable;
see \sec{sec:lang}.
There is also a large literature on non-linear effects in light
propagation, starting with a paper by Zel'dovich in 1964 \cite{Zeldovich:1964}
(see \cite{Rasanen:2008a} for further discussion and references).
However, the question phrased above has rarely been the focus
of light propagation studies, and it has not received a definitive
answer.

We assume that metric perturbations remain small, and
concentrate on the second issue.
The expansion rate and the distance both involve second derivatives
of the metric, like the density perturbation.
The question is not whether it is possible to have large local
deviations: the variation of the local expansion rate between
different regions is of order unity in the real universe.
The issue is whether it follows from the smallness of metric
perturbations that the distribution of the expansion rate is
such that the fluctuations cancel in the average, and whether
corrections to the distance and redshift are correspondingly small.

We show that if the spacetime is close to FRW in the sense that
metric perturbations, their time derivatives and first
spatial derivatives are small and if the observer
four-velocity is close to its background value and close
to the four-velocity which defines the hypersurface of
averaging, the redshift and the average expansion rate
remain close to the FRW case.
Such a result does not hold for the angular diameter distance.

In \sec{sec:setup} we set up the formalism and state our assumptions and
in \sec{sec:light} we show that the change in the redshift is small
and explain why this is not the case for the angular diameter distance.
In \sec{sec:exp} we consider the average expansion rate.
In \sec{sec:disc} we discuss our results, in particular the relation
to Newtonian gravity, and in \sec{sec:sum} we summarise the situation.

\section{The spacetime geometry} \label{sec:setup}

\subsection{Kinematics and the equation of motion}

\para{Two frames.}

We mostly follow the notation of \cite{Rasanen:2009b};
for reviews of the covariant formalism, see
\cite{Ehlers:1961, Ellis:1971, Ellis:1998c, Tsagas:2007}.
We denote the four-velocity of the observers by $u^\a$.
We consider a spacelike hypersurface $\sN$ and denote
the unit vector orthogonal to $\sN$ by $n^\a$.
We will take averages on this hypersurface;
at this stage, $\sN$ is completely general.
Both vectors are normalised to unity,
$u_\a u^\a=n_\a n^\a=-1$. The tensors which project on the
hypersurface orthogonal to $n^\a$ and the rest space orthogonal
to $u^\a$ are, respectively,
\bea \label{h}
  h_{\a\b} &\equiv& g_{\a\b} + n_\a n_\b \el
  h_{\a\b}^{(u)} &\equiv& g_{\a\b} + u_\a u_\b \ ,
\eea

\noindent and we denote by $\path_\a$ the spatial covariant
derivative which is completely projected on $\sN$, e.g.
$\path_\b f_\a = h_{\b}^{\ \d} h_{\a}^{\ \c} \nabla_\d f_\c$.
Without loss of generality, we write
\bea \label{n}
  u^\a = \gamma ( n^\a + v^\a ) \ ,
\eea

\noindent where $v_\a n^\a=0$ and $\gamma=-n_\a u^\a=(1-v^2)^{-1/2}$
with $v^2\equiv v_\a v^\a$.

It is useful to decompose the gradient of $n_\a$ as
\bea \label{gradn}
  \nabla_\b n_\a
  &=& \frac{1}{3} h_{\a\b} \theta + \sigma_{\a\b} - A_\a n_\b \ ,
\eea

\noindent where $\theta\equiv\nabla_\a n^\a$ is the volume expansion rate, $\sigma_{\a\b}\equiv\path_{(\b} n_{\a)}-\frac{1}{3} h_{\a\b} \theta$ is the shear tensor and $A_\a\equiv n^\b\nabla_\b n_\a$ is the acceleration vector.
The analogous decomposition of the gradient of $u_\a$ is
\bea \label{gradu}
  \nabla_\b u_\a
  &=& \frac{1}{3} h^{(u)}_{\a\b} \theta^{(u)} + \sigma^{(u)}_{\a\b} + \omega^{(u)}_{\a\b} - A^{(u)}_\a u_\b \ ,
\eea

\noindent where $\omega^{(u)}_{\a\b}\equiv\nabla_{[\b} u_{\a]}+A^{(u)}_{[\a} u_{\b]}$
is the vorticity tensor and the other quantities are defined in
the same manner as those in \re{gradn}.

We assume that the relation between the geometry and the matter
content is given by the Einstein equation (we use units in which
$8\pi\GN=1$, $\GN$ being Newton's constant), 
\bea \label{Einstein}
  G_{\a\b} &=& T_{\a\b} = \rho u_\a u_\b + p h^{(u)}_{\a\b} + 2 q_{(\a} u_{\b)} + \pi_{\a\b} \ ,
\eea

\noindent where we have without loss of generality
decomposed the energy-momentum tensor $T_{\a\b}$ with respect to $u^\a$.
Here $\rho\equiv u^\a u^\b T_{\a\b}$ is the energy density,
$p\equiv\frac{1}{3} h^{\a\b} T_{\a\b}$ is the pressure,
$q_\a\equiv - h_\a^{\ \b} u^\c T_{\b\c}$ is the energy flux and
$\pi_{\a\b}\equiv h_{\a}^{\ \c} h_{\b}^{\ \d} T_{\c\d} - \frac{1}{3} h_{\a\b} h^{\c\d} T_{\c\d}$
is the anisotropic stress.

\subsection{The near-FRW assumption}

\para{The metric.}

We write the metric as
\bea \label{metric}
  \rmd s^2 = - (1 + 2 \Phi) \rmd \tb^2 + 2 \alpha_i \rmd\tb\rmd x^i +  \left( [1 - 2 \Psi] f_{ij} + \chi_{ij} \right) a(\tb)^2 \rmd x^i \rmd x^j \ ,
\eea

\noindent where
$f_{ij}\equiv (1+K\delta_{kl} x^k x^l/4)^{-2}\delta_{ij}\equiv f \delta_{ij}$
is the metric of a three-dimensional homogeneous and isotropic
space with constant curvature $6 K/a^2$.
The scale factor is normalised to unity today, $a(\tb_0)=1$.
We have $\delta^{ij}\chi_{ij}=0$.
We define $H\equiv\adot/a$, where dot means derivative
with respect to the coordinate time $\tb$.
This form of the metric is completely general.
We refer to the spacetime obtained when
$\Phi=\Psi=0, \a_i=0, \chi_{ij}=0$ as the background, and
refer to these functions as perturbations.
We choose the Poisson gauge, which is defined by
$\delta^{ij}\alpha_{i|j}=0$, $\delta^{jk}\chi_{ij|k}=0$,
where $|$ indicates covariant derivative with respect to $f_{ij}$.

We want the spacetime to be close to FRW and the coordinate
system to be close to the coordinates where the background
looks homogeneous and isotropic, so
we assume that the metric perturbations are small.
For the scalar functions $\Phi$ and $\Psi$,
we can simply demand that their values are small everywhere.
The magnitude of $\alpha_i$ and $\chi_{ij}$ depends on the
coordinate system, so we have to be a bit more careful.
If the background space is negatively curved, $f$ diverges
at $r=2/\sqrt{-K}$, and $f$ approaches zero as $r$ goes
to infinity for either negative or positive spatial curvature.
Correspondingly, if a field $A^i$ has finite norm with
regard to the background space, $\bar{g}_{ij}A^iA^j=a^2f\delta_{ij}A^iA^j$
(where $\bar{g}_{ij}$ is the spatial background metric),
the components $A^i$ will vanish or diverge as $f$ diverges or
vanishes, respectively.
We define the background-normalised spatial components of any
field as $A^\hi\equiv a\sqrt{f} A^i$ and $A_\hi\equiv (a\sqrt{f})^{-1} A_i$
(and correspondingly for fields with more than one spatial
index) to avoid this coordinate divergence.
The requirement that the perturbations are small can now
be stated as
$\e(x)\equiv\max(|\Phi|, |\Psi|, |\alpha_\hi|, |\chi_{\hi\hj}|)\ll1$.
For $\alpha_i$ and $\chi_{ij}$ we can equivalently say
$\delta^{ij} \alpha_\hi \alpha_\hj=\bar{g}^{ij} \alpha_i\alpha_j\lesssim\e^2, \delta^{ik} \delta^{jl} \chi_{\hi\hj} \chi_{\hat{k} \hat{l}}=\bar{g}^{ik} \bar{g}^{jl} \chi_{ij} \chi_{kl}\lesssim\e^2$.
In summary, we assume that the spacetime metric is everywhere
perturbatively near the same global background.
We further assume that the background spatial
curvature is not significantly larger than the background expansion
rate, $|K|/a^2\lesssim H^2$, i.e. we do not consider near-static
spacetimes.

We are interested in modes whose wavelengths are not long,
that is to say modes for which first spatial derivatives are
large compared to (or of the same order as) the perturbations,
but still smaller than unity, given $a|H|$ as the
comparison scale, $1\gg|\pt_\hi\e|\gtrsim\e$,
where we have defined
$\pt_\hi\equiv (a\sqrt{f})^{-1}\pat_i/|H|$.\footnote{If the background
expansion is expanding and decelerating, this condition becomes
stronger over time, since $1/(a|H|)$ increases.
Conversely, in an accelerating expanding background the
condition becomes weaker.
For a collapsing background the situation is reversed.
This is assuming that the time-dependence of $\e$ does
not overcome that of $a|H|$.}
We also assume that
time evolution is slow compared to spatial changes, more precisely
that $|\dot\e|\lesssim |H| \e$, i.e. time derivatives are
at most of the same order of magnitude as the background time scale.
We make no assumptions about second derivatives of the
metric perturbations, they can be comparable to the background
quantities or larger.

\para{The Einstein tensor.}

The components of the Einstein tensor for the near-FRW metric \re{metric} are
\bea
  \label{G00} G_{00} &\simeq& 3 H^2 + 3 \frac{K}{a^2} + 2 \Psi^{|k}_{\ \ |k} + \sO(\e\pat^2\e, \pat\e\pat\e, \bar\Gamma\e\pat\e, \pat\bar\Gamma\e, \bar\Gamma^2\e, \bar\Gamma H\e, H^2\e) \\
  \label{G0i} G_{0\hi} &\simeq& - \ha \alpha_{\hi\ \, |k}^{\,\, |k} + \frac{K}{2 f^{\ha} a^2} x^k \pat_k \alpha_\hi + 2 \pat_\hi (\Psidot + H\Phi) + \sO(\e\pat^2\e, \pat\bar\Gamma\e, \bar\Gamma^2\e, \bar\Gamma H\e, H^2\e) \\
  \label{Gij} G_{\hi\hj} &\simeq& - \left( 2 \acc + H^2 + \frac{K}{a^2} \right) \d_{ij} + (\Psi-\Phi)^{|i}_{\ \ |j} - (\Psi-\Phi)^{|k}_{\ \ |k} \d_{ij} - \pat_{(\hi} \dot\alpha_{\hj)} - 2 H \pat_{(\hi} \alpha_{\hj)} \el
  && - \frac{1}{2} a^2 \chi_{\hi\hj\ \, |k}^{\ \, |k} + \frac{K}{f^{\ha}} x^k \pat_k \chi_{\hi\hj} + \sO(\e\pat^2\e, \pat\e\pat\e, \bar\Gamma\e\pat\e, \pat\bar\Gamma\e, \bar\Gamma^2\e, \bar\Gamma H\e, H^2\e) \ ,
\eea

\noindent where $\simeq$ indicates dropping subleading terms in metric
perturbations and their derivatives; in the remainder terms
we have not kept track of the indices. We use the symbol $\bar\Gamma$
to refer to the background spatial Christoffel symbols;
$\pat$ indicates $\pat_\hi$ and $\pat^2$ indicates a combination
of two spatial derivatives (likewise for $\pt$ and $\pt^2$).
To simplify the bookkeeping, we take in what follows
$\bar\Gamma\lesssim|H|$, in line with the assumption
that we do not consider near-static universes.

Because the Einstein equation is second order, there
are at most two derivatives acting on a metric perturbation, so the
structure remains close to linear theory. Note that this not an
expansion in powers of the metric perturbation: in that case
perturbations and their derivatives would be considered to be of the same
order \cite{Bruni:1996}. When derivatives are large, this is
inconsistent, as the first and second derivatives of the metric
perturbations are effectively new expansion parameters
\cite{Rasanen:2010a} (see also \cite{Green:2010}).

\para{The four-velocity.}

Observables such as the redshift, the angular diameter
distance and the local expansion rate depend
on the observer four-velocity $u^\a$.
According to observations, deviations of galaxies
from the mean flow are small over large scales.
These local deviations often go by the name peculiar
velocities. In linear theory and in the Poisson gauge,
it is simple to identify $u^i$ as the physical
velocity around the mean flow.
However, defining the peculiar velocity in a more
general context and translating the observational
constraint into a well-defined mathematical statement
is not straightforward \cite{peculiar, Tsagas:2007}.

The difference between the actual value of $u^\a$ and
its background value is gauge-dependent, so the physical
meaning of it being small is not obvious.
(Requiring metric perturbations to remain small is open
to the same criticism.)
For example, it is always possible to adopt the comoving gauge
where $u^i=0$, though then metric perturbations become large at
the same time as density perturbations.
On the other hand, $|u^\hi|\sim1$ does not necessarily
contradict any observations, any more than
metric perturbations of order unity do.
The physical peculiar velocity would need to be defined with
respect to a physically defined velocity field describing
the mean flow.
We will simply look at the difference from the background
in the Poisson gauge, like we do with the metric perturbations.
For the background we have $\bar{u}^\a=\delta^{\alpha0}$, so
we are interested in whether the conditions $|u^0-1|,|u^\hi|\ll1$ hold.
Given the normalisation $g_{\a\b} u^\a u^\b=-1$ and the smallness
of metric perturbations, the first condition follows from the second,
so we only need to check whether $|u^\hi|\ll1$.
From the $0i$ component of \re{Einstein}, we have
\bea \label{ui}
  u_\hi = \frac{ G_{0\hi} - \a_\hi p - u_0 q_\hi - \pi_{0\hi} }{ (\rho+p) u_0 + q_0 } \ ,
\eea

\noindent The relation \re{ui} shows what is required in terms
of metric perturbations and the matter content to keep $u_\hi$ small.
According to \re{G0i} the leading contribution
to $G_{0\hi}$ is $\alpha_{\hi\ \, |k}^{\,\, |k}\simeq (a\sqrt{f})^{-3/2}\nabla^2\alpha_i$,
so smallness of metric perturbations and their first derivatives
is not enough to guarantee that $u_\hi$ would remain small\footnote{It
is important that time derivatives of metric perturbations
remain small, otherwise $u_\hi$ is in general large, as
$G_{0\hi}$ always involves second derivatives of the metric.}.
We could make the additional assumption
$|\alpha_{\hi\ \, |k}^{\,\, |k}/H^2|\lesssim|\pt\e|$
for the metric, and assume for the matter content that
$|q_\hi/(\rho+p)|\lesssim|\pt\e|$,
$|\pi_{0\hi}/(\rho+p)|\lesssim|\pt\e|$,
and that there is no negative pressure so large that
we would have $|\rho+p|\ll\rho$.
(Note that the magnitude of $\pi_{\hi\hj}$ is unconstrained.)
Under these conditions, \re{ui} gives $|u_\hi|\ll1$.
These conditions are sufficient, but not necessary, as there
can be cancellations among the different terms in \re{ui}.
We will therefore simply assume that $u_\hi\sim\sO(\pt\e)$
without specifying which of these conditions hold.

If the observer motion is geodesic, $A^{(u)}_\a=0$, it follows
from the geodesic equation that $|u_\hi|\sim|\pt\e|\ll1$.
In particular, this is the case if the matter is dust
(as viewed by the observer).
The condition $|\alpha_{\hi\ \, |k}^{\,\, |k}/H^2|\ll1$ then
also follows automatically.

\section{Light propagation} \label{sec:light}

\subsection{The redshift} \label{sec:red}

Most cosmological observations probe redshifts and distances.
Let us first consider the redshift. 
The redshift measured by the observer is given by the change
in photon energy between emission and observation,
$1+z=E_\mathrm{e}/E_\mathrm{o}$.
In the geometrical optics approximation, light travels on
null geodesics \cite{Schneider:1992} (page 93), \cite{Sasaki:1993},
and the energy is
\bea \label{E}
  E &=& - u_\a k^\a \el
  &\simeq& k^0 [ 1 + \sO(\pt\e) ] \ ,
\eea

\noindent where $k^\a$ is the photon momentum, tangent to 
a null geodesic. It is useful to split $k^\a$ as
\bea
  k^\a = E ( u^\a + e^\a ) \ ,
\eea

\noindent where $u_\a e^\a=0$, $e_\a e^\a=1$.
We define $\frac{\rmd}{\rmd\eta}\equiv (u^\a+e^\a)\pat_\a$.
The component $k^0$ is determined by the null geodesic equation
\bea \label{nullgeo}
  0 &=& k^\a \nabla_\a k^0 \el
  &=& k^\a \pat_\a k^0 + \Gamma^0_{\a\b} k^\a k^\b \el 
  &\simeq& k^\a \pat_\a k^0 + H f_{ij} k^i k^j + \sO(H k^0 k^0 \pt\e) \el
  &\simeq& k^\a \pat_\a k^0 + H k^0 k^0 + \sO(H k^0 k^0 \pt\e) \ ,
\eea

\noindent where we have on the last line used the null condition
$g_{\a\b} k^\a k^\b=0$. From \re{E} and \re{nullgeo} we have
\bea \label{z}
  1 + z &\simeq& \exp{ \left( \int_\mathrm{e}^\mathrm{o}\rmd\eta \left[ H + \sO(H\pt\e) \right] \right) } \el
  &\simeq& \left(\frac{a_\mathrm{e}}{a_\mathrm{o}}\right)^{-1} [ 1 + \sO(\pt\e) ] \ .
\eea

\noindent The redshift is to first approximation given
by the inverse of the background scale factor.
In other words, as long as metric perturbations are small
(and the other assumptions hold), emission which is
nearly isotropic at the source looks nearly isotropic
to the observer.
The converse is not true: near-isotropy of the redshift
of the cosmic microwave background does not imply that
the metric would be close to FRW \cite{Rasanen:2009a}.

\subsection{The distance}

While the relation between the redshift and the background scale
factor remains to leading order unchanged from the FRW case,
we cannot say whether changes in the redshift are small
or large unless we know how the background scale factor is
related to observables.
More generally, the redshift is only observationally meaningful
if expressed in relation to other observable quantities, such
as the angular diameter distance or the age of the universe.
In particular, the redshift-distance relation can change
significantly, because the change in the angular diameter
distance $D_A$ can be large\footnote{Recall that the
luminosity distance $D_L$ is related to
the angular diameter distance via $D_L=(1+z)^2 D_A$
in a general spacetime \cite{Ellis:1971},
\cite{Schneider:1992} (page 111), \cite{Etherington:1933}.}.
The reason why the redshift remains close to its background value
is that the photon momentum is given by a first order differential
equation where first derivatives of the metric enter via the
Christoffel symbols, and second derivatives do not make an
appearance. In contrast, the equation for the angular diameter
distance is second order. We have
\bea \label{DAeq}
 \frac{\rmd^2 D_A}{\rmd\l^2} &=& - \left[ 4 \pi\GN \big( \rho + p - 2 q_\a e^\a + \pi_{\a\b} e^\a e^\b \big) E^2 + \sigmat^2 \right] D_A \ ,
\eea

\noindent where $\frac{\rmd}{\rmd\l}\equiv k^\a \nabla_\a$ and $\sigmat^2$
is the null shear scalar; see
\cite{Rasanen:2009b, Schneider:1992, Sasaki:1993} for details.

When converting the derivative with respect to the affine parameter
$\lambda$ to derivative with respect to the observable
redshift, we have\footnote{In general, this change of variables does not make sense, because the redshift is not always monotonic along the null geodesic, so there is no function $D_A(z)$ \cite{Rasanen:2008b, Rasanen:2009b}.}
\bea \label{ltoz}
  \frac{\rmd D_A}{\rmd\l} &=& \pat_z D_A \frac{\rmd z}{\rmd\l} \el
  &=& \pat_z D_A E ( u^\a \pat_\a + e^\a \pat_\a ) z \el
  &\simeq& \pat_z D_A E \left[ \pat_0 z + e^i\pat_i z + \sO(\pt\e \pat_\a z) \right] \ .
\eea

\noindent Because the redshift receives corrections of order
$\pt\e$, the conversion factor \re{ltoz} involves second
derivatives of the metric.
While the perturbations do not substantially change the redshift,
they change the relation between the redshift and the affine
parameter.
This corresponds to changing the local expansion rate, shear
and/or acceleration \cite{Rasanen:2008b, Rasanen:2009b}.

The work \cite{Enqvist:2009} provides an example where
metric perturbations around a matter-dominated spatially flat FRW
background are small, and their time derivatives and 
first spatial derivatives are also small, and the four-velocity
perturbation is small, but the angular diameter distance is
very different from the background, and is designed to exactly
reproduce the best-fit $\Lambda$CDM FRW model.
The model studied in \cite{Enqvist:2009} is spherically symmetric.
If the universe is statistically homogeneous and isotropic (and has
a finite homogeneity scale) and the distribution evolves slowly,
it can be argued that the change due
to the spatial derivatives in \re{ltoz} cancels in the integrals along
the null geodesic over distances longer than the homogeneity scale.
In that case the distance is for typical light rays
to leading order determined by the average energy density, pressure and
expansion rate \cite{Rasanen:2008b, Rasanen:2009b}. If the pressure
can be neglected, the angular diameter distance is determined by
the average expansion rate and the value of the average density
today. If the average expansion rate is close to the background,
the distance is expected to be close to the FRW case, in agreement
with Swiss Cheese studies of light propagation \cite{cheese}.
The argument should be studied in more detail and made more
rigorous.

An alternative to the integral approach considered here is
to expand $D_A$ as a series in $z$ (leaving aside that in
the real universe there is no function $D_A(z)$), or
vice versa \cite{Kristian:1966, Clarkson:2000}, as
recently discussed in \cite{Clarkson:2011}.
It would seem that significant variations
in different directions in the distance are expected
when second derivatives of the perturbations are large.
Such an expansion is only useful for small redshifts
or distances, and the cancellations for the distance are
expected to occur only over large scales, so the two
pictures are not in contradiction.

\section{The expansion rate} \label{sec:exp}

\subsection{The local expansion rate} \label{sec:local}

In addition to the redshift and the distance, we can observe
the expansion rate. Let us now consider the average expansion
rate, its relation to the background scale factor $a$ and
the effect of perturbations on the evolution of $a$.
The volume expansion rate measured by the observer is
\bea \label{theta}
  \theta^{(u)} &=& \nabla_\a u^\a \el
  &=& \pat_\a u^\a + \Gamma^\a_{\a\b} u^\b \el
  &\simeq& 3 H + \pat_i u^i + \sO(H\pt\e) \ ,
\eea

\noindent where we on the last line applied the metric \re{metric}.
As $\pat_i u^i\sim\sO(H\pt^2\e)$, the local expansion rate can
have large variations in different regions.
However, the presence of large local variations does not necessarily
mean that the average expansion rate would change significantly; that
depends on the distribution of the fluctuations.

We assume that the observers are moving non-relativistically with
respect to the averaging frame, $v\ll1$;
to simplify the bookkeeping, we assume that $v\lesssim\sO(\pt\e)$.
When considering averages, it is useful to decompose vectors and
tensors in the direction orthogonal to and directions parallel to
the averaging hypersurface $\sN$, instead of the background
time and space directions.
To this end, we split $n^\a$ into a vector $m^\a$ whose gradient
gives (approximately) the background expansion rate $3 H$ and a
vector $p^\a$ which lies along $\sN$.
We define the former by setting $m^\a=\delta^{\a0}$ in the
coordinates \re{metric}, and the latter by
\bea \label{t}
  p^\a \equiv n^\a + \frac{m^\a}{m^\b n_\b} \ ,
\eea

\noindent with the components
$p^0\simeq\sO(\e, \pt\e\pt\e)$,
$p^i=n^i\simeq\sO(\pt\e)$.
With these definitions, we have
\bea \label{thetadec}
  \theta^{(u)} &=& g^{\a\b} \nabla_\b u_\a \el
  &=& \gamma ( \path_\a n^\a + \path_\a v^\a + A_\a v^\a ) + \gamma^3 ( v^\a v^\b \path_\b v_\a + v^\a n^\b \nabla_\b v_\a ) \el
  &\simeq& \path_\a n^\a + \path_\a v^\a + \sO(H\pt\e\pt\e\pt^2\e) \el
  &=& - \frac{1}{m^\b n_\b} \path_\a m^\a + \frac{1}{(m^\b n_\b)^2} m^\a \path_\a (m^\c n_\c) + \path_\a ( p^\a + v^\a ) + \sO(H\pt\e\pt\e\pt^2\e) \el
  &\simeq& 3 H + \path_\a s^\a + \sO(H\pt\e\pt\e\pt^2\e) \ ,
\eea

\noindent where we have defined $s^\a\equiv p^a + v^\a$.
Like $p^\a$ and $v^\a$ (but unlike $u^i$), $s^\a$ is a vector along $\sN$.
For the shear we have similarly
\bea
  \sigma^{(u)}_{\a\b} &=& h^{(u)}_{\a (\mu} h^{(u)}_{\nu)\b} \nabla^\nu u^\mu - \frac{1}{3} h^{(u)}_{\a\b} \theta^{(u)} \el
  &\simeq& \path_\b s_\a - \frac{1}{3} h_{\a\b} \path_\c s^\c + \sO(H\pt\e\pt^2\e) \ .
\eea

\subsection{The average expansion rate} \label{sec:average}

As we want to average $\theta^{(u)}$ over $\sN$, we need the
relation between the background time $\tb$ and the time which
is constant on $\sN$, which we denote by $t$ (note that unless
$A^\a=0$, $t$ is not a proper time). We have
\bea
  \pat_t \tb &=& n^\a \pat_\a\tb \el
  &=& n^0 \el
  &\simeq& 1 + \sO(\pt\e\pt\e)
\eea

\noindent and $\path_\a\tb\sim\sO(\pt\e)$,
so the difference between the times $\tb$ and $t$ is small.
Nevertheless, the difference in the volume element between the
hypersurfaces of constant $\tb$ and constant $t$ can be large.
If the spatial coordinates differ by $\sO(H^{-1}\pt\e)$, the
Jacobian of the coordinate transformation is $\sO(\pt^2\e)$,
which is of the same order as the density perturbations.
(For dust this is rather obvious: because mass is conserved,
the density is inversely proportional to the volume element.)
For an explicit example in a case where the hypersurface of averaging is
taken to be the hypersurface of constant proper time measured by
observers, see \cite{Kolb:2004}.
The average of the expansion rate \re{thetadec} on $\sN$ is
\bea \label{avtheta}
  \av{\theta^{(u)}} &\simeq& \av{ 3 H(\tb) + \path_\a s^\a + \sO(H\pt\e\pt\e\pt^2\e) } \el
  &\simeq& 3 H(t) + \av{ \path_\a s^\a } + \sO(H\pt\e\pt\e\pt^2\e) \el
  &\simeq& 3 H(t) + \sO(H \pt\e/(HL), H\pt\e\pt\e\pt^2\e) \ ,
\eea

\noindent where $\av{\path_\a s^\a}$ reduces to a boundary term
which is suppressed by $\sO(\pat\e)$ and enhanced by $1/(HL)$,
where $L^3$ is the proper volume of the averaging region.
As long as the region is not so much smaller than the background
Hubble scale that it would overcome the smallness of
$s^\a\sim\sO(\pt\e)$, the boundary term remains subdominant.
There can be large local fluctuations, but they cancel over large
volumes.
For this argument it is crucial that $\path_\a$ is a
derivative along $\sN$ and $s^\a$ is a vector on $\sN$.
The assumption that time derivatives of perturbations are
not large is also essential; without it, all large
second derivative terms would not reduce to suppressed
boundary terms.

If the distribution on $\sN$ is statistically homogeneous
and isotropic and the averaging region is at least as large
as the homogeneity scale, any total derivative would
give a small contribution, even if the amplitude of the
vector field were not small. The reason is that a total
derivative corresponds to flux through the boundary, and without
a preferred direction this should be equal in both
directions across the boundary, up to statistical fluctuations
\cite{Notari:2005, Rasanen:2008a, Rasanen:2009b}.
Boundary terms vanish identically for periodic boundary
conditions, which are used in simulations and implicitly assumed
in Fourier series decomposition.

We have established that the average expansion rate is close
to the background quantity $H$. The averaging hypersurface $\sN$
has been kept general, up to the condition that the difference
between the four-velocity $n^\a$ orthogonal to $\sN$
and the observer four-velocity $u^\a$ is small.
In general, different hypersurfaces of averaging
give different results \cite{Geshnizjani, Rasanen:2004},
and relevant averages are those which give an approximate
description of observable quantities.
Arguments about cancellations in integrals
related to the redshift and the distance indicate
that these are the averages taken on the hypersurface
of statistical homogeneity and isotropy
\cite{Rasanen:2006b, Rasanen:2008a, Rasanen:2008b, Rasanen:2009b}.
However, we see that varying the choice of
hypersurface does not change the leading order result as long
as the difference between the two frames is non-relativistic
\cite{Rasanen:2009b}.
It was argued in \cite{Rasanen:2009b} that the observationally
relevant expansion rate is $\nabla_\a n^\a$, which is related
to the hypersurface of statistical homogeneity and isotropy,
while we have considered $\nabla_\a u^\a$.
Nevertheless, their averages are close, because the difference
between the two frames is non-relativistic.

\subsection{The background expansion rate} \label{sec:back}

We have established that the redshift and the expansion
rate are given in terms of the background scale factor
in the same way as in FRW universes.
However, this does not necessarily mean that their relation to time
would be the same as in the FRW case, because the evolution
of the scale factor $a$ (or equivalently the background expansion
rate $H=\adot/a$) could be different.
In usual perturbation theory, equations are split up in powers of
the metric perturbations, and equations at each order are assumed to be
satisfied separately \cite{Bruni:1996}. In particular, the evolution
of background quantities is taken to be independent of the
perturbations. However, this is an extra assumption which does
not follow from the equations of motion.
At linear order, the equations for background quantities are
the same as in the FRW case (as long as the average of the linear
perturbations vanishes).
Beyond linear order, the average of the perturbations does not
vanish, and when derivatives of the perturbations become large,
higher order terms could have a significant impact on the average.

Let us see what happens in the present case, when second
derivatives of the perturbations can be even larger than the
background quantities, and we do not assume that the equations
are satisfied order by order.
From \re{Einstein} and \re{G00}--\re{Gij}, we have
\bea
  \label{rho} \rho &\simeq& 3 H^2 + 3 \frac{K}{a^2} + 2 \Psi^{|k}_{\ \ |k} + \sO(\pt\e\nabla^2\alpha_\hi, \pt\e\pt\e\pat^2\e, \e\pat^2\e, \bar\pat\Gamma\e) \\
  \label{p} - p &\simeq& 2 \acc + H^2 + \frac{K}{a^2} + \frac{2}{3} (\Psi-\Phi)^{|k}_{\ \ |k} + \sO(\pt\e\nabla^2\alpha_\hi, \pt\e\pt\e\pat^2\e, \e\pat^2\e, \pat\bar\Gamma\e) \ .
\eea

Let us first average \re{rho} on $\sN$.
The first two terms depend only on the background time $\tb$,
which is close to the time $t$, so we get simply
$\av{3 H(\tb)^2 + 3 \frac{K}{a(\tb)^2}}\simeq 3 H(t)^2 + 3 \frac{K}{a(t)^2} + \sO(H^2\pt\e)$.
For the third term we have
\bea
  2 \av{\Psi^{|k}_{\ \ |k}} &\simeq& 2 \av{\path_\a \path^\a \Psi} + \sO(\pt\e\pt\e \pat^2\e) \ .
\eea

\noindent This is a total derivative of a vector that has a small
amplitude, so it is suppressed on the same grounds as before.
The average of \re{p} is analogous, and we obtain
\bea
  \label{rho2} && \!\!\!\!\!\!\!\!\!\!\!\!\!\!\!\!\!\!\!\!\! \av{\rho} \simeq 3 H(t)^2 + 3 \frac{K}{a(t)^2} + \sO(H\pat\e/(HL), \pt\e\nabla^2\alpha_\hi, \pt\e\pt\e\pat^2\e, \e\pat^2\e, \pat\bar\Gamma\e)  \\
  \label{p2} && \!\!\!\!\!\!\!\!\!\!\!\!\!\!\!\!\!\!\!\!\! \av{p} \simeq - 2 \frac{\partial_t^2 a(t)}{a(t)} - H(t)^2 - \frac{K}{a(t)^2} + \sO(H\pat\e/(HL), \pt\e\nabla^2\alpha_\hi, \pt\e\pt\e\pat^2\e, \e\pat^2\e, \pat\bar\Gamma\e) \ .
\eea

\noindent In other words, to leading order the evolution
of $a$ is given by the FRW equations (note that here $a$ and $H$
are functions of the physical time $t$, not the background time $\tb$).
The average expansion rate and the redshift are therefore
related to the time $t$ in the same way as the background
quantities are related to $\tb$, up to small corrections.

In \cite{Rasanen:2010a}, it was argued that the magnitude of the
corrections to the average expansion rate cannot be resolved in
usual perturbation theory once the density perturbations become
non-linear. The reason was that when the expansion rate is written
as a series in powers of the metric perturbation, the contribution
of higher order terms is not suppressed, and the series expansion
becomes useless when second derivatives of the perturbations
become large.
The feature of the present treatment which makes it possible to
establish the amplitude of the corrections is that second derivatives
of the perturbations are not treated perturbatively.
The infinite series discussed in section 2.1 of \cite{Rasanen:2010a}
arises from expanding the denominator of $u_\hi$ in \re{ui} in a
power series, though this is not obvious in a perturbative treatment.
Here we utilise the feature that the average of $\pat_i u^i$
reduces (approximately) to a boundary term at any order
in perturbation theory. This fact has not been recognised in
previous perturbation theory studies
\cite{Futamase, Wetterich:2001, Rasanen, Notari:2005, Kolb:2005a, Kolb:2005b, Rasanen:2010a, Seljak, Russ:1996, Kolb:2004, Siegel:2005, Ishibashi:2005, Baumann:2010, Clarkson:2011}
(for further references and discussion, see \cite{Rasanen:2010a}).

\section{Discussion} \label{sec:disc}

\subsection{The Buchert equations}

The Buchert equations show the effect of deviations from
homogeneity and isotropy on the average expansion rate
in general terms \cite{Buchert:1995, Buchert:1999, Buchert:2001}.
Let us see how the above result for the average expansion
rate emerges from them.
We consider the expansion rate $\theta=\nabla_\a n^\a$;
as noted above, the difference between the average of $\theta$
and the average of $\theta^{(u)}$ is small. The average of the expansion
rate $\theta$ evolves according to the equations \cite{Rasanen:2009b}
\bea
  \label{Ray} \pat_t \av{\theta} + \frac{1}{3} \av{\theta}^2 &\simeq& - \ha \left( \av{\rho} + 3 \av{p} \right) + \sQ \el
  && + \sO(\pt\e\pat^2\e/(LH), \pt\e\nabla^2\a_\hi, \pt\e\pt\e\pt^2\e\pat^2\e) \\
  \label{Ham} \frac{1}{3} \av{\theta}^2 &\simeq& \av{\rho} - \frac{1}{2}\av{\sR} - \frac{1}{2}\sQ + \sO(\pt\e\pt\e\pt^2\e\pat^2\e) \\
  \label{cons} \pat_t \av{\rho} + \av{\theta} ( \av{\rho} + \av{p} ) &\simeq& - \av{\theta p} + \av{\theta} \av{p} - \av{\sigma_{\a\b} \pi^{\a\b}} \el
  && + \sO(H \nabla^2\a_\hi/(LH), \pat\e\nabla^2\a_\hi, H^3\pt\e\pt\e\pt^2\e\pt^2\e) \ ,
\eea

\noindent where we have taken into account that $A^\a\sim\sO(\pat\e)$
and that the related time dilation (i.e. deviation of $t$ from proper time)
is small. (Note that $\nabla^2\a_\hi$ makes an appearance as a
boundary term, which is suppressed for large averaging volumes
if we assume statistical homogeneity and isotropy, or if we assume
that $|q_\hi|\lesssim H^2$.)
Here $\sR$ is the spatial curvature scalar on $\sN$, and
the backreaction variable $\sQ$ is
\bea \label{Q}
  \sQ &\equiv& \frac{2}{3}\left( \av{\theta^2} - \av{\theta}^2 \right) - \av{\sigma_{\a\b}\sigma^{\a\b}} \el
  &=& \av{\path_\a n^\a \path_\b n^\b - \path_\b n_\a \path^\b n^\a } - \frac{2}{3} \av{\path_\a n^\a}^2 \el
  &\simeq& \av{\path_\a p^\a \path_\b p^\b - \path_\b p_\a \path^\b p^\a } - \frac{2}{3} \av{\path_\a p^\a}^2 + \sO(\pt\e\pt^2\e\pat^2\e) \el
  &=& \av{ \path_\a ( p^\a \path_\b p^\b - p^\b \path_\b p^\a ) + p^\b [\path_\a, \path_\b] p^\a } - \frac{2}{3} \av{\path_\a p^\a}^2 + \sO(\pt\e\pt^2\e\pat^2\e)  \el
  &=& \av{\path_\a ( p^\a \path_\b p^\b - p^\b \path_\b p^\a )} - \av{\sR_{\a\b} p^\a p^\b} - \frac{2}{3} \av{\path_\a p^\a}^2 + \sO(\pt\e\pt^2\e\pat^2\e) \ ,
\eea

\noindent where we have on the second line used the definitions of
$\theta$ and $\sigma_{\a\b}$ given in \re{gradn}, and $\sR_{\a\b}$ is
the spatial curvature tensor. (We have also used the fact that to
leading order $\path_{\b} p_{\a}=\path_{(\b} p_{\a)}$.)
The boundary terms are small for the
same reasons as before and the spatial curvature
contribution is suppressed by two powers of $p^\a\sim\sO(\pt\e)$.
This structure where $\sQ$ is almost a boundary term is close
to the Newtonian case, as we discuss in \sec{sec:Newton}.

The integrability condition between \re{Ray} and \re{Ham} reads
\bea \label{int}
  && \pat_t {\av{\sR}} + \frac{2}{3} \av{\theta} \av{\sR} = - \pat_t \sQ - 2 \av{\theta} \sQ - 2 \av{\theta p} + 2 \av{\theta} \av{p} - 2 \av{\sigma_{\a\b} \pi^{\a\b}} \el
  && + \sO(H\nabla^2\a_\hi/(LH), \pat\e\pat^2\e/(LH), \pat\e\nabla^2\a_\hi, H^3\pt\e\pt\e\pt^2\e\pt^2\e)  \ ,
\eea

\noindent so if $\sQ$ is small and the other terms are small
(which would have to be looked at separately), the average
spatial curvature evolves in the same manner as in the FRW
case. The local spatial curvature scalar is
\bea \label{R}
  \!\!\!\!\!\!\!\!\!\!\!\! \sR &=& 2 G_{\a\b} n^\a n^\b - \frac{2}{3} \theta^2 + \sigma_{\a\b} \sigma^{\a\b} \el
  &\simeq& 6 \frac{K}{a^2} + 4 \Psi^{|k}_{\ \ |k} - 4 H \path_\a p^\a - \path_\a ( p^\a \path_\b p^\b - p^\b \path_\b p^\a ) + \sO(\pt\e\pt^2\e\pat^2\e) \el
  &\simeq& 6 \frac{K}{a^2} + 4 \frac{1}{a^2 f} \nabla^2 \Psi - 4 H \path_\a p^\a - \path_\a ( p^\a \path_\b p^\b - p^\b \path_\b p^\a ) + \sO(\pt\e\pt^2\e \pat^2\e) \ .
\eea
 
\noindent When the density contrast is non-linear, there are
typically large local variations in the spatial curvature,
like in the expansion rate. In the average, these
large deviations cancel up to boundary terms, and the
leading behaviour is the same as in the FRW case,
$\av{\sR}\simeq 6 K/a(t)^2 + \sO(\pt\e\pt^2\e\pat^2\e)$.
If $\Psi$ is constant in time, the time-dependence of the
first term in \re{R} is also $a^{-2}$, and it can be viewed
as a ``renormalisation'' of the background spatial curvature
constant $K$ \cite{Kolb:2005a}. In general, $\Psi$ depends on
time, and such an interpretation is not valid.

\subsection{The redshift and the average expansion rate}

In a general spacetime, the redshift measured by an observer is
\bea \label{zgen}
  1+z &=& \exp\left( \int_\mathrm{e}^\mathrm{o}\rmd\eta \left[ \frac{1}{3} \theta^{(u)} + A^{(u)}_\a e^\a + \sigma^{(u)}_{\a\b} e^\a e^\b \right] \right) \ .
\eea

\noindent It might appear that the change in the redshift due to perturbations
would be of the order of the change in the average expansion rate.
As we have seen, the latter reduces to a boundary term which, while
small for sufficiently large regions, may be important for small
domains. However, according to \sec{sec:red}, the change in redshift
is always small under our assumptions, irrespective of the distance
travelled by the light. Let us see how these facts are reconciled. We have
\bea
  && \int_\mathrm{e}^\mathrm{o}\rmd\eta \left( \frac{1}{3} \theta^{(u)} + A^{(u)}_\a e^\a + \sigma^{(u)}_{\a\b} e^\a e^\b \right) \el
&\simeq& \int_\mathrm{e}^\mathrm{o}\rmd\eta \left[ H + \frac{1}{3} \path_\a s^\a + u^\b \nabla_\b u_\a e^\a + \left( \path_\b s_\a - \frac{1}{3} h_{\a\b} \path_\c s^\c \right) e^\a e^\b + \sO(H\pt\e\pt^2\e) \right] \el
  &\simeq& \ln\frac{a_\mathrm{o} }{a_\mathrm{e}} + \int_\mathrm{e}^\mathrm{o}\rmd\eta \, \left[ \path_\b s_\a e^\a e^\b + \sO(H\pt\e\pt^2\e) \right] \el
  &=& \ln\frac{a_\mathrm{o} }{a_\mathrm{e}} + \int_\mathrm{e}^\mathrm{o}\rmd\eta \, \left[ e^\b \path_\b (s_\a e^\a) - s_\a e^\b \path_\b e^\a + \sO(H\pt\e\pt^2\e) \right] \el
  &=& \ln\frac{a_\mathrm{o} }{a_\mathrm{e}} + \int_\mathrm{e}^\mathrm{o}\rmd\eta \, \left[ \frac{\rmd (s_\a e^\a)}{\rmd\eta} - u^\b \nabla_\b (s_\a e^\a) + \sO(H\pt\e\pt^2\e) \right] \el
  &\simeq& \ln\frac{a_\mathrm{o} }{a_\mathrm{e}} + \Bigg|_\mathrm{e}^\mathrm{o} s_\a e^\a + \sO(\pt\e\pt^2\e) \el
  &\simeq& \ln\frac{a_\mathrm{o} }{a_\mathrm{e}} + \sO(\pt\e\pt^2\e) \ .
\eea

The leading order deviation in the local expansion rate cancels with
a term in the projected shear. The remaining part
$e^\a e^\b \path_\b s_\a$ is locally large, but at leading order it reduces
to a total derivative in $\eta$ and thus to a small boundary term.
In \cite{Rasanen:2008b} it was noted that while the cancellation
(up to a boundary term) between the expansion rate and the shear
in $\sQ$ can be understood in terms of the Newtonian limit, it
is not clear whether their cancellation in the redshift could be
understood in a similar manner.
We now see that it follows from the smallness of $u_\hi$
and the smallness of perturbations of the Christoffel symbols.

It is only the sum of the contributions of $\theta-3H$ and
$\sigma^{(u)}_{\a\b} e^\a e^\b$ which is small, not either term
individually.
In a statistically homogeneous and isotropic space where
the distribution evolves slowly, the integral of
$\sigma^{(u)}_{\a\b} e^\a e^\b$ alone should be strongly
suppressed for typical light rays over long distances
(because $\sigma^{(u)}_{\a\b}$ has no preferred directions
while $e^\a$ varies only slowly) \cite{Rasanen:2008b, Rasanen:2009b}\footnote{To
be precise, the argument should be formulated in the $n^\a$ frame
in terms of the decomposition \re{gradn} of $\nabla_\b n_\a$
\cite{Rasanen:2009b}.}.
It then follows that the contribution of $\theta-3H$ is also small, in
agreement with the argument that the contribution of the expansion rate
is given by the spatial average (which is close to $3H$) if the space
is statistically homogeneous and isotropic and the distribution
evolves slowly \cite{Rasanen:2008b, Rasanen:2009b}.

If matter consists of discrete clumps instead of a continuous
fluid, it has been argued that there could be a large effect on
the redshift \cite{Clifton}.
Given the above results, this would imply that the spacetime
cannot be written in terms of a near-FRW metric and a small $u_\hi$,
or that the geometrical optics approximation is not valid \cite{Rasanen:2009b}

\subsection{Relation to Newtonian gravity} \label{sec:Newton}

In Newtonian cosmology, the Raychaudhuri equation \re{Ray}
is identical to its general relativity counterpart
\cite{Ellis:1971, Ellis:1990}.
In contrast, the counterpart of the
Hamiltonian constraint \re{Ham} emerges only as the
first integral of the Raychaudhuri equation, whereas
in general relativity it is an independent equation.
This difference corresponds to the absence of spatial
curvature in Newtonian gravity \cite{Ellis:1971}.
This is related to the fact that there are no covariant derivatives,
only ordinary derivatives, which commute; as a result, the
backreaction variable $\sQ$ contains only boundary terms \cite{Buchert:1995}.
When the system is isolated, i.e. boundary terms vanish,
the first integral of \re{Ray} gives \re{Ham} with $\sQ=0$ and
a conserved energy term proportional to $a^{-2}$.
For this reason the evolution of the scale factor in the Newtonian theory
is always the same as in the FRW case, regardless of the
amplitude of perturbations: in particular, accelerating expansion
due to inhomogeneities is not possible\footnote{In Newtonian
gravity, it is also impossible to get acceleration by introducing exotic
matter with negative pressure, because pressure does not gravitate.}.

In contrast, in general relativity the Hamiltonian constraint
\re{Ham} involves the average spatial curvature term, which can
have non-trivial evolution
\cite{Rasanen:2005, Rasanen:2008a, Buchert:2008, Rasanen:2010a, newrevs}.
However, if perturbations of the Christoffel symbols are small,
the spatial structure remains close to Newtonian theory
(apart from possible background curvature).
Absence of spatial curvature is related to the Newtonian
constraint that the magnetic component of the Weyl tensor vanishes,
$H_{\a\b}=0$
\cite{Ellis:1971, Ellis:1994, Kofman:1995, Matarrese:1995, Ehlers:2009}.
In general relativity, the magnetic part of the Weyl tensor
decomposed with respect to $n^\a$ is
\bea
  H_{\a\b} &=& \e_{\c\d(\a} \path^\c \sigma^\d_{\ \b)} \ ,
\eea

\noindent where $\e_{\a\b\c}\equiv\eta_{\a\b\c\d} n^\d$ is the volume
element on $\sN$, with $\eta_{\a\b\c\d}$ being the spacetime volume
element. In the Newtonian limit, the shear can be written as
$\sigma_{\a\b}=\path_\a\path_\b\phi-\frac{1}{3} h_{\a\b} \path_\c
\path^\c \phi$, where $\phi$ is a scalar function identified with
the gravitational potential \cite{Ellis:1971, Kofman:1995, Matarrese:1995}.
We then have
\bea
  H_{\a\b} &=& \e_{\c\d(\a} \path^\c \path^\d \path_{\b)} \phi = \ha \e_{\c\d(\a} [\path^\c, \path^\d] \path_{\b)} \phi = \ha \e_{\c\d(\a} \sR^{\c\d \,\ \e}_{\ \ \b)} \path_\e \phi \ .
\eea

\noindent The three-dimensional Riemann tensor $\sR_{\a\b\c\d}$
vanishes if and only if $\sR_{\a\b}$ does, because the Weyl tensor
is zero in three dimensions.
This relates the absence of spatial curvature and backreaction
in Newtonian gravity to the lack of propagating degrees of
freedom\footnote{In general relativity, the magnetic
component $H_{\a\b}$ and the electric component $E_{\a\b}$ have coupled
evolution equations, which have wave solutions.
In the Newtonian theory, $H_{\a\b}$ is zero and
$E_{\a\b}$ does not have an evolution equation.}.
For the metric \re{metric} and our approximation
of treating metric perturbations and their first derivatives as
small, $\Phi, \Psi$ and $\chi_{ij}$ and their derivatives do not
contribute to $H_{\a\b}$ at leading order, while second
derivatives of $\alpha_i$ do enter and take the system
far from the Newtonian behaviour, as also happens with $G_{0i}$.
(In general relativity, for an irrotational dust fluid,
$H_{\a\b}$ is trivially zero in usual perturbation theory at
first order, implying that the theory has a linearisation
instability \cite{silent}; see also \cite{Death:1976}.
We have not assumed that the matter is dust.)

In the usual post-Newtonian formalism \cite{Will:1981} (page 86), \cite{PN}
it is assumed that $\Phi, \Psi\sim\e$, $\alpha_i\sim\e^{3/2}$
and $\chi_{ij}\sim\e^2$.
If in the present context we were to similarly assume that
$\alpha_i$ is smaller than $\Phi$ and $\Psi$ so that its
second derivatives are small, the situation would be closer
to the usual post-Newtonian formulation.
If we start with only scalar perturbations at linear
order and solve the equations of motion order by order, then at second
order we have $\alpha_\hi\sim\e\pt\e$ and $\chi_{ij}\sim\e^2$
\cite{Matarrese:1997}. However, it is not
clear whether this hierarchy persists once second derivatives of
perturbations become large and the equations cannot be solved
order by order.
In the post-Newtonian formalism, each time derivative further reduces
the order of magnitude by $\e$, whereas spatial derivatives do
not change it. In contrast, in the present approach we assume that
time derivatives are (at most) of the same order of magnitude as the
background, while spatial derivatives increase the order of magnitude.
(The post-Newtonian formalism is constructed around Minkowski space,
so the background does not involve a scale.)

It might seem promising to study backreaction in a
post-Newtonian approximation.
However, an essential feature of the usual post-Newtonian
scheme is that the system is finite and isolated, which is
not the case in cosmology.
In fact, Newtonian gravity has a well-defined initial
value problem only for isolated systems, periodic boundary conditions
or fractal distributions with vanishing mean density 
\cite{Ellis:1994, Ehlers:1996, Ehlers:1999, Norton:1999, Szekeres, Gabrielli:2010}.
Related to this, the non-relativistic limit of taking the
speed of light to infinity is singular, so solutions of
the limiting Newtonian equations are in general not limits
of solutions of the relativistic equations \cite{Szekeres, Senovilla:1997}.
Nevertheless, there has been work on post-Newtonian formulations
of cosmology
\cite{Tomita, Szekeres, Matarrese:1995, Matarrese, Puetzfeld},
all in contexts where metric perturbations are assumed to remain small.
A comparison of Newtonian cosmological simulations and relativistic
analytical treatment was made in \cite{Alonso:2010} for a specific
spherically symmetric dust configuration. Good agreement was found
between the two theories in this highly symmetric case.

\subsection{Local and global backgrounds} \label{sec:lang}

As long as the metric and the four-velocity remain close to FRW,
there is no significant backreaction (with the caveats we have
mentioned).
The Christoffel symbols are given by first derivatives
of the metric, and we assume that first derivatives of
perturbations are small.
Therefore powers of the perturbed Christoffel symbols higher than
the first are negligible, and the structure remains close to linear theory.
Even though variation in the Riemann tensor can be large, the
geodesic equation involves only the Christoffel symbols, so
the effect of curvature is locally small for light propagation
and for timelike geodesic motion.

The result does not imply that in order for backreaction to
be significant, there would have to be large local deviations in the
Christoffel symbols. It simply means that all regions should not
be close to the same global background \cite{Ellis:2011}.
Metric perturbations and their first spatial derivatives can
still remain small with regard to a local background, which
is different in different regions.
With reference to a global background, the metric perturbations
or their first spatial derivatives in some regions would then be large.

For example, consider a stabilised region with a constant
non-zero density, such as a dark matter halo, in an expanding
spatially flat dust background.
From \re{rho} it follows that $\nabla^2\Psi$ has a decaying
part proportional to $a^{-1}$ (corresponding to the falling
background expansion rate) and a growing a part proportional to $a^2$
(corresponding to the constant density).
In terms of the background coordinate time $\tb$, the perturbation
$\Psi$ will become larger than unity as the universe expands.
This does not mean that there would be locally strong gravitational
effects: metric perturbations remain small (away from regions of
high mass concentration) if the metric is expanded around
Minkowski space in the local region.
It is only the difference between the global background and
the appropriate local background which is diverging.
Of course, the evolution of the metric perturbations should be
determined in detail from the equations of motion, and
this argument only shows that the metric perturbation $\Psi$
of a stabilised region grows initially like $a^2$ if the
perturbations and their derivatives are initially small.
There is a complication in that it would be more appropriate
to consider the proper time measured by observers (or at least
the time $t$ which is constant on $\sN$) and not the unphysical
background time $\tb$. Even if the two times are near each
other, the difference in the corresponding spatial derivatives is of
order unity once second derivatives of the metric become large. It is
therefore not straightforward to extract the proper time dependence
of $\Psi$ from the background time dependence of $\nabla^2\Psi$.

Possibly ordinary perturbation could be used as a null test: by
calculating quantities such as the variance of the expansion rate
\cite{Li}\footnote{The study \cite{Li} was mistakenly criticised
in \cite{Rasanen:2008a, Rasanen:2010a} for misestimating the magnitude
of the boundary term. However, the analysis is still constrained by the
applicability of second order perturbation theory.}
in perturbation theory and comparing to observations
we might try to rule out the assumption that perturbation
theory holds. However, to obtain predictions, it is usually not
enough to know that the amplitude of the metric perturbations
is small, their evolution has to be known as well.

Straightforward perturbation theory with regard to the metric is not
suited to the case when metric perturbations are not small around
a global background, but are instead small only with regard to
local backgrounds which are different in different regions.
A full analytical treatment is unfeasible, but it might be
possible to obtain a simplified system of equations which
could be treated via statistical methods \cite{Rasanen:2008a, peakrevs}
or numerical simulations. One possibility could be to use the
covariant formulation of the evolution and constraint equations
\cite{Ehlers:1961, Ellis:1971, Ellis:1998c, Tsagas:2007},
which deals directly with physical degrees of freedom, so that the
problem can be discussed without assumptions about perturbativity
of the metric around some background. However, it is not clear whether
there is a tractable approximation which would include the cosmologically
relevant degrees of freedom.

Let us emphasise how the present treatment differs from previous work
\cite{Futamase, Seljak, Russ:1996, Wetterich:2001, Rasanen, Kolb:2004, Siegel:2005, Ishibashi:2005, Rasanen:2010a, Baumann:2010}
(see \cite{Rasanen:2010a} for more references and discussion).
Most previous studies apply either usual or post-Newtonian
perturbation theory, where all quantities are expanded
in a series of either the metric perturbation or
peculiar velocity, and solved order by order.
However, when second derivatives of metric perturbations
become large in cosmology, this procedure is inconsistent,
because second order quantities can be larger than first
order quantities, $\pt^2\e\pt^2\e\gg\e$.
In contrast, we have not assumed that second derivatives
of metric perturbations are small or split the quantities
order by order.
Therefore our result applies beyond usual perturbation theory.
Also, we have defined the averaging hypersurface using physical
criteria, taken the volume element into account, and correctly
identified the local expansion rate measured by the observer,
unlike in some previous work.
We have also studied the observable redshift and angular diameter
distance directly.

Recently, an interesting approach has been introduced
to tackle the backreaction problem without having
to assume that second derivatives are small \cite{Green:2010}.
The idea is to consider a family of metrics $g_{\a\b}(\l)$ which
depend on some small parameter $\lambda$. The background is identified
as the metric $g_{\a\b}(0)$, and perturbation theory is developed
in terms of $\l$. A novelty of the formalism is that space and time
derivatives of the metric are not assumed to be analytic in $\l$,
and they have a well-defined limit as $\l\rightarrow0$ only when
smeared over a local region of the background spacetime.
It is then shown (with some other assumptions)
that if the metric is close to the background, the background
satisfies the Einstein equation with only an additional effective
radiation term, and corrections due to perturbations are small.
The formalism is interesting in that the notion of a homogeneity
scale is incorporated into the analysis in a natural manner,
and the calculations are mathematically rigorous.
However, the connection to the real universe is somewhat
unclear. The physical interpretation of the parameter
$\l$ and the adopted scaling of various quantities
with $\l$ is not obvious. In the present case,
we use standard general relativity, and there are no
extra assumptions to be made.
(However, in \cite{Green:2010}, it is not assumed
that the first derivatives of metric perturbations would be small.)
Note that it is important to consider correctly defined
observables, as smoothing and calculating observables do
not in general commute: in \cite{Green:2010},
the redshift and the angular diameter distance are
not considered.
The cautionary example \cite{Enqvist:2009} shows that small
metric perturbations and small peculiar velocities do not
guarantee that changes to the angular diameter distance are small.
Also, as noted in \cite{Ellis:2011}, the local smoothing
considered in \cite{Green:2010} is a different procedure
from averaging over large scales. (The smoothing in
\cite{Green:2010} is done with respect to the background
space, not the physical space.)
To the extent one can make a comparison, the results of
\cite{Green:2010} and the present work do not appear to be
in disagreement.
Both studies share the weakness that their starting point
is that the spacetime is close to FRW, an assumption which
should be carefully looked at.

\section{Conclusion} \label{sec:sum}

%\para{Summary.}

It has been claimed that the effect of non-linear structures on
the average expansion rate and on light propagation is small
in the real universe, because the metric remains close
to FRW at all times.
The argument has two parts: that the metric remains close to FRW, and that
this implies that the change in the average expansion rate and light
propagation is small.
We have considered the second part of the argument. The
relevant quantities depend not only on the metric, but also on
its first and second derivatives.
Second derivatives of metric perturbations have variations
of order unity after structures become non-linear, so smallness
of the metric perturbations alone is not a sufficient condition
for backreaction to be small.
We have made the assumptions that time derivatives
and first spatial derivatives of metric perturbations are small
and the perturbation of the observer four-velocity is small.
(The last assumption implies certain conditions on second
derivatives of the metric and/or the matter content.)
It then follows that the redshift is to leading order given
by the background scale factor.
We further assume that the difference between the observer
four-velocity and the four-velocity which defines the hypersurface
of averaging is small.
The difference between the average expansion rate and the background
expansion rate then reduces to a boundary term which is small as
long as the averaging domain is not much smaller than the Hubble scale,
and the background expansion rate evolves in the same manner
as in the FRW case.
This can be understood from the fact that perturbations of
the Christoffel symbols remain small, so the structure is close
to Newtonian cosmology, where backreaction reduces to a
boundary term \cite{Buchert:1995}.

However, even with these assumptions, perturbations can have
a large effect on the angular diameter distance, as
demonstrated in \cite{Enqvist:2009}.
It has been argued that if the space is statistically
homogeneous and isotropic and the distribution evolves slowly,
then the distance is determined by the average expansion rate,
and the change in the distance is small, too
\cite{Rasanen:2008b, Rasanen:2009b}.
The issue should be studied in more detail.

The assumptions needed for the proof show that
smallness of metric perturbations and their time
derivatives and first spatial derivatives
is not sufficient for the effect on the redshift and
the average expansion rate to be small.
The assumption about the smallness of the
deviation of the observer four-velocity $u^\a$ from the
background is crucial.
In general, the deviation of $u^\a$ from the background is a
gauge-dependent quantity which cannot be straightforwardly
identified as the deviation from the physical mean flow
determined from observations.
Nevertheless, if we assume that the observer moves along
a geodesic (which is the realistic case in the late universe),
the smallness of the deviation of $u^\a$ follows from the
assumptions about metric perturbations and their derivatives.
It would be useful to have a definition of the
peculiar velocity that would be valid in a general cosmological
spacetime \cite{peculiar} and that would correspond to the
observational use of the term, so that the observed smallness
of this quantity could be used as an input.

If backreaction is significant, its effect cannot be expressed
in terms of a changed FRW background nor small perturbations
around a FRW universe
\cite{Rasanen:2006b, Rasanen:2008a, Rasanen:2008b, Rasanen:2009b, Rasanen:2010a, newrevs}.
Rather, perturbations can remain small only with respect to a
local region, and if we insist on a global background metric
instead of a patchwork of backgrounds, then metric perturbations
or perturbations of the four-velocity have to be large.
The important issue is the behaviour of physical quantities,
not in which form the metric or four-velocity can be written.
In the real universe, fluctuations of the Riemann tensor
and the expansion rate are not small: local variations
of the expansion rate are of the same order as the
measured deviation of the average expansion rate
from the matter-dominated FRW value.
The question is whether the distribution of local values
is such that variations cancel when considering
the average expansion rate, the redshift and the distance.
Further focusing on perturbativity,
perhaps in the context of a patchwork of regions, might be a
useful way towards reliably quantifying backreaction.

\acknowledgments

I thank Carl Bender, Stephen Green, Hans M\"{u}hlen,
Dominik Schwarz, Robert Wald and Clifford Will for useful discussions
and Washington University in St Louis, the University of Chicago,
LAPTH and the Pettersson Institute for Theoretical Physicists for
their hospitality.\\

%McGill University, Niels Bohr International Academy, Dark Cosmology Centre,
%King's College London and Bielefeld University 

\end{document}